\newcommand{\CLASSINPUTtoptextmargin}{19mm}%
\newcommand{\CLASSINPUTbottomtextmargin}{43mm}%
\newcommand{\CLASSINPUTinnersidemargin}{12.9mm}%
\newcommand{\CLASSINPUToutersidemargin}{12.9mm}%
\documentclass[conference,10pt,a4paper]{IEEEtran}
\usepackage{amsmath}
\usepackage{times}
\usepackage{graphicx}
\usepackage{multirow}
\usepackage[none]{hyphenat}
\usepackage{float}
\usepackage{subfig}
\usepackage[binary-units]{siunitx}
\usepackage{import}
\usepackage{glossaries}
\usepackage{color}
\DeclareSIUnit{\sample}{S}

\newacronym{bbif}{BBIF}{baseband \& intermediate frequency}
\newacronym{soc}{SoC}{systems-on-chip}
\makeatletter

\def\@maketitle{\newpage
\bgroup\par\addvspace{0.5\baselineskip}\centering%
\ifCLASSOPTIONtechnote
   {\bfseries\large\@IEEEcompsoconly{\sffamily}\@title\par}\vskip 1.3em{\lineskip .5em\@IEEEcompsoconly{\sffamily}\@author
   \@IEEEspecialpapernotice\par{\@IEEEcompsoconly{\vskip 1.5em\relax
   \@IEEEtitleabstractindextextbox{\@IEEEtitleabstractindextext}\par
   \hfill\@IEEEcompsocdiamondline\hfill\hbox{}\par}}}\relax
\else
   \vskip0.2em{\EuMWtitlesize\ifCLASSOPTIONtransmag\bfseries\LARGE\fi\@IEEEcompsoconly{\sffamily}\@IEEEcompsocconfonly{\normalfont\normalsize\vskip 2\@IEEEnormalsizeunitybaselineskip
   \bfseries\Large}\@title\par}\vskip1.0em\par
   \ifCLASSOPTIONconference%
      {\@IEEEspecialpapernotice\mbox{}\vskip\@IEEEauthorblockconfadjspace%
       \mbox{}\hfill\begin{@IEEEauthorhalign}\@author\end{@IEEEauthorhalign}\hfill\mbox{}\par}\relax
   \else
      \ifCLASSOPTIONpeerreviewca
         {\@IEEEcompsoconly{\sffamily}\@IEEEspecialpapernotice\mbox{}\vskip\@IEEEauthorblockconfadjspace%
          \mbox{}\hfill\begin{@IEEEauthorhalign}\@author\end{@IEEEauthorhalign}\hfill\mbox{}\par
          {\@IEEEcompsoconly{\vskip 1.5em\relax
           \@IEEEtitleabstractindextextbox{\@IEEEtitleabstractindextext}\par\hfill
           \@IEEEcompsocdiamondline\hfill\hbox{}\par}}}\relax
      \else
         \ifCLASSOPTIONtransmag
           {\@IEEEspecialpapernotice\mbox{}\vskip\@IEEEauthorblockconfadjspace%
            \mbox{}\hfill\begin{@IEEEauthorhalign}\@author\end{@IEEEauthorhalign}\hfill\mbox{}\par
           {\vspace{0.5\baselineskip}\relax\@IEEEtitleabstractindextextbox{\@IEEEtitleabstractindextext}\vspace{-1\baselineskip}\par}}\relax
         \else
           {\lineskip.5em\@IEEEcompsoconly{\sffamily}\sublargesize\@author\@IEEEspecialpapernotice\par
           {\@IEEEcompsoconly{\vskip 1.5em\relax
            \@IEEEtitleabstractindextextbox{\@IEEEtitleabstractindextext}\par\hfill
            \@IEEEcompsocdiamondline\hfill\hbox{}\par}}}\relax
         \fi
      \fi
   \fi
\fi\par\addvspace{0.0\baselineskip}\egroup}

\def\EuMWtitlesize{\@setfontsize{\EuMWtitlesize}{24}{24pt}}
\def\EuMWauthorsize{\@setfontsize{\EuMWauthorsize}{11}{11pt}}
\def\EuMWaffilsize{\@setfontsize{\EuMWaffilsize}{10}{10pt}}
\def\EuMWcaptionsize{\@setfontsize{\EuMWcaptionsize}{9}{10pt}}
\def\EuMWbibsize{\@setfontsize{\EuMWbibsize}{8}{10pt}}

\def\@IEEEauthorblockNstyle{\EuMWauthorsize\@IEEEcompsocnotconfonly{\sffamily}\@IEEEcompsocconfonly{\large}}
\def\@IEEEauthorblockAstyle{\EuMWaffilsize\@IEEEcompsocnotconfonly{\sffamily}\@IEEEcompsocconfonly{\itshape}\@IEEEcompsocconfonly{\large}}
\def\@IEEEauthordefaulttextstyle{\EuMWauthorsize\@IEEEcompsocnotconfonly{\sffamily}\sublargesize}

\def\thebibliography#1{\section*{\refname}%
    \addcontentsline{toc}{section}{\refname}%
    \EuMWbibsize\@IEEEcompsocconfonly{\small}\vskip 0.3\baselineskip plus 0.1\baselineskip minus 0.1\baselineskip
    \list{\@biblabel{\@arabic\c@enumiv}}%
    {\settowidth\labelwidth{\@biblabel{#1}}%
    \leftmargin\labelwidth
    \advance\leftmargin\labelsep\relax
    \itemsep \IEEEbibitemsep\relax
    \usecounter{enumiv}%
    \let\p@enumiv\@empty
    \renewcommand\theenumiv{\@arabic\c@enumiv}}%
    \let\@IEEElatexbibitem\bibitem%
    \def\bibitem{\@IEEEbibitemprefix\@IEEElatexbibitem}%
\def\newblock{\hskip .11em plus .33em minus .07em}%
\ifCLASSOPTIONtechnote\sloppy\clubpenalty4000\widowpenalty4000\interlinepenalty100%
\else\sloppy\clubpenalty4000\widowpenalty4000\interlinepenalty500\fi%
    \sfcode`\.=1000\relax}

\long\def\@makecaption#1#2{%
\ifx\@captype\@IEEEtablestring%
\par\@IEEEtabletopskipstrut
\else
\@IEEEfigurecaptionsepspace
\fi
\setbox\@tempboxa\hbox{\normalfont\footnotesize {#1.}\nobreakspace\nobreakspace #2}%
\ifdim \wd\@tempboxa >\hsize%
\setbox\@tempboxa\hbox{\normalfont\footnotesize {#1.}\nobreakspace\nobreakspace}%
\parbox[t]{\hsize}{\normalfont\footnotesize\noindent\unhbox\@tempboxa#2}%
\else
\ifCLASSOPTIONconference \hbox to\hsize{\normalfont\footnotesize\hfil\box\@tempboxa\hfil}%
\else \hbox to\hsize{\normalfont\footnotesize\box\@tempboxa\hfil}%
\fi\fi
\ifx\@captype\@IEEEtablestring%
\@IEEEtablecaptionsepspace
\else
\fi}

\newlength\tablecaptiontotableskip
\newlength\figuretocaptionskip
\def\@IEEEfigurecaptionsepspace{\vskip\figuretocaptionskip\relax}%
\def\@IEEEtablecaptionsepspace{\vskip\tablecaptiontotableskip\relax}%

\def\abstract{\normalfont%
\@IEEEabskeysecsize\bfseries\textit{\abstractname}\,\bfseries\textit{---}\,%
\@IEEEgobbleleadPARNLSP}%

\def\IEEEkeywords{\normalfont%
\@IEEEabskeysecsize\bfseries\textit{\IEEEkeywordsname}\,\bfseries\textit{---}\,%
\@IEEEgobbleleadPARNLSP}%
\def\endIEEEkeywords{\relax\vspace{0.67ex}%
\par\if@twocolumn\else\endquotation\fi%
\normalsize\normalfont}%

\DeclareRobustCommand*{\EuMWauthorrefmark}[1]{\raisebox{0pt}[0pt][0pt]{\textsuperscript{\footnotesize{#1}}}}%
\def\@IEEEauthorblockNtopspace{0ex}
\def\@IEEEauthorblockAtopspace{1mm}
\def\tablename{Table}
\def\thetable{\arabic{table}}
\def\IEEEkeywordsname{Keywords}
\def\subsubsection{\@startsection{subsubsection}{3}{\z@}{1.5ex plus 1.5ex minus 0.5ex}%
{0.7ex plus .5ex minus 0ex}{\normalfont\normalsize\itshape}}%
\newlength{\CPheadmatchindent}%
\def\@seccntformat#1{\hbox to\CPheadmatchindent{\csname the#1dis\endcsname}\hskip 0.1em \relax}
\IEEEilabelindentA \parindent
\IEEEilabelindent \IEEEilabelindentA
\IEEEelabelindent \parindent
\IEEEdlabelindent \parindent
\IEEElabelindent \parindent
\makeatother

\begin{document}
\raggedbottom
%
%
%
\title{A Radar Kit for Hands-On Distance-Learning}
%
%
\author{%
\IEEEauthorblockN{%
Markus Gardill\EuMWauthorrefmark{\#1}, Tushar Tandon\EuMWauthorrefmark{*2}
}
\IEEEauthorblockA{%
    \EuMWauthorrefmark{\#}Chair of Electronic Systems and Sensors, Brandenburgische Technische Universität Cottbus-Senftenberg, Germany\\
    \EuMWauthorrefmark{*}Computer Science VII - Robotics and Telematics, Julius-Maximilians-Universität Würzburg (JMU), Germany\\
    \EuMWauthorrefmark{1}markus.gardill.de@ieee.org, \EuMWauthorrefmark{2}tushar.tandon@uni-wuerzburg.de\\
}
}
%
\maketitle
%
%
\begin{abstract}
We present an approach to experimental radar systems education based on a combination of commercial low-cost hardware with modern open-source software technologies.
Following a discussion of the general top-level architecture of flexible, software-defined radar systems, we introduce the specific selection of subsystems, their capabilities, and current system limitations.
Compared to existing approaches to practical radar education, a more top-level modular design with a greater focus on performance and flexibility of baseband processing is selected while reducing the complexity of circuit and subsystem assembly and total system cost.
We present example measurements obtained from the radar kit.
The radar kit allows for bringing a radar lab to the students instead of students into the labs.
It enables practical hands-on radar education also in distance-only-learning scenarios.
\end{abstract}
\begin{IEEEkeywords}
education, fmcw, radar
\end{IEEEkeywords}
%
%

\section{Introduction}

Using a ``practice first'' approach to RF electronic systems' education was the original idea behind the famous  `coffee-can radar' (CCR), developed at the  Massachusetts Institute of Technology (MIT) Lincoln Laboratory \cite{charvat2012}.
Many educators across different universities adapted it for practical electronics engineering education \cite{hernandez-jayo2015}.
Continuous work on improving this educational platform is reported \cite{brinkmann2019, kolodziej2019}.
When transforming the classroom-based courses utilizing the CCR to an online course, several challenges were identified \cite{kolodziej2019}, and a major redesign of the hardware was necessary to simplify the construction and operation of the radar at the student's home.
However, it still is recommended to have ``soldering iron, Dremel tool, ... as well as a multimeter and oscilloscope to help test the radar'' at hand \cite{mit_build_a_radar_course_website}, which might not be available at each student's home.

Although its users can perform advanced measurements like, e.g., synthetic aperture radar (SAR) \cite{kolodziej2019}, the primary motivation of the CCR has not been radar education. 
Naturally, the system performance from a radar perspective is limited: the operating frequency is $f_{\text{c}} = \SI{2.4}{\giga\hertz}$ and a maximum bandwidth of $W_{\text{RF}} = \SI{80}{\mega\hertz}$ limits range resolution.
The intermediate-frequency (IF) path is limited to $W_{\text{IF}} = \SI{15}{\kilo\hertz}$, allowing for data acquisition and processing using a low-cost microcontroller  \cite{kolodziej2019}.

Lab hardware developed explicitly for radar experimentation typically targets higher performance goals than the original CCR.
In \cite{kurniawan2017}, the authors addressed especially real-time processing capabilities as well as RF bandwidth limitation of the CCR and designed a variant with increased RF bandwidth and sampling rate.
A \gls{soc} development board combining ARM processor cores with field-programmable gate array (FPGA) logic and data acquisition at a rate of $f_{\text{s}} = \SI{1}{\mega\sample\per\second}$ is added to the original CCR design in \cite{brinkmann2019}.
However, both approaches rely on the original CCR RF front end concept.
They thus cannot meet the performance of radar lab platforms such as the one proposed in \cite{diewald2018}, where an educational two-transmit, eight-receive antenna radar system operating at the $\SI{24}{\giga\hertz}$ industrial, scientific, and medical (ISM) bands is designed.
However, this combination of high-performance radar front end and a commercial multi-channel data-acquisition and control platform does not focus on bringing the setup to the student’s home.
Commercial development platforms such as the INRAS RadarBook \cite{inras_radar_book} indeed offer standalone solutions for radar systems research, but their comparably high cost and complexity typically do not target the educational sector.

\begin{figure}
    \includegraphics[width=\columnwidth]{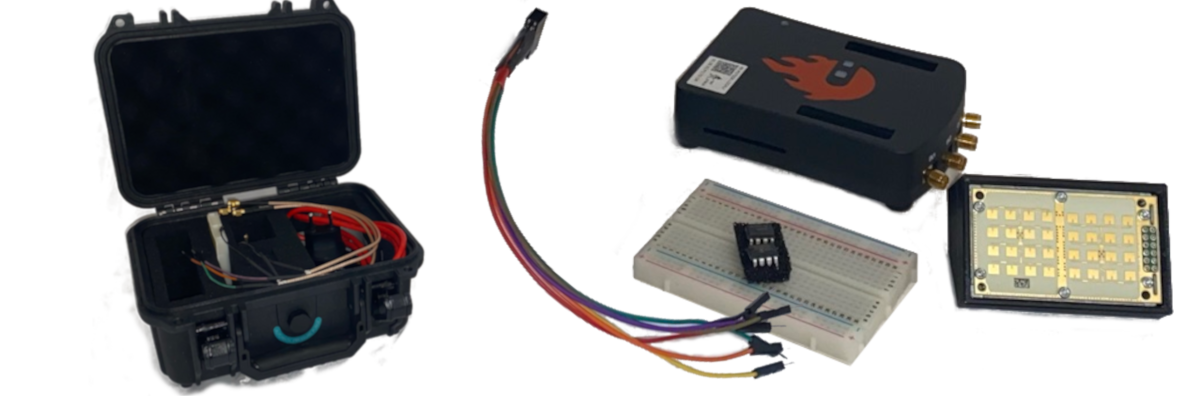}
    \caption{The radar kit is a completely self-contained experimental radar platform in a rugged case for easy and safe transportation.}
    \label{fig:lab_kit_photo}
\end{figure}

The authors of this contribution were looking for an educational radar platform with a performance similar to modern industrial radars, e.g., operation in the \SI{24}{\giga\hertz} bands with RF bandwidths over \SI{100}{\mega\hertz}, supporting various types of modulated continuous-wave (CW) radar waveforms.
Additional requirements for being deployed in distance-learning courses include ease of implementation, complete self-contained hardware, open-source software, and low cost.
Since no suitable solution was identified, the authors designed a new approach for such an educational platform.
It consists of just two commercial off-the-shelf (COTS) main hardware building blocks and has a total cost of around 300,- € per kit.
Platform-independent open-source software allows students to use their own Linux, Windows or Mac computers with the system, see fig. \ref{fig:lab_kit_photo}.
In the remainder of this paper, the ideas of that radar kit are shared.
Further information, software, and the course materials are available on GitHub \cite{drmarkusg2021}.

\section{Main Building Blocks}

\begin{figure}
    \small
    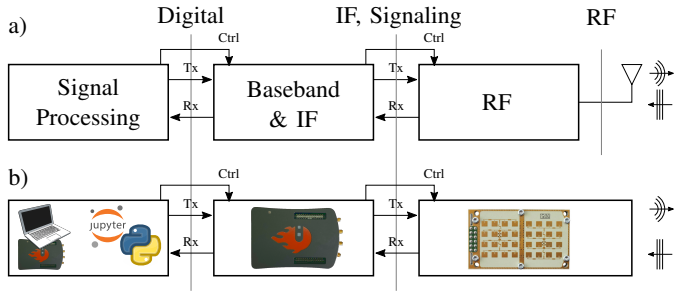
    \caption{a) Top-level architecture of a radar system divided into four main building blocks with three types of interfaces and b) the hardware subsystems utilized in the radar kit.}
    \label{fig:radar_system_top_level_architecture}
\end{figure}

A radar system can be subdivided into four functional building blocks, compare fig. \ref{fig:radar_system_top_level_architecture} a).
Their function slightly differs if used in transmit (Tx) or receive (Rx) direction.

The signal processing section defines a digital representation of the used radar waveform in Tx direction.
In Rx direction, it extracts information about the radar scene from analyzing the waveform obtained from the \gls{bbif} stage with the knowledge of the Tx waveform.
It further might create additional digital control information to set the radar system operational state, e.g., the setting of variable-gain amplifiers or the (de-)activation of radar resources, e.g. Tx channels.
The \gls{bbif} stage converts the digital representation of the radar Tx waveform into its analog equivalent.
It further converts the digital control information into analog control signals interfacing to the RF front end.
In Rx direction, it acquires the analog output of the RF front end and outputs a digital representation of that waveform.
The RF front end generates the RF signal to be radiated based on the analog representation of the waveform obtained from the \gls{bbif} stage.
In Rx direction, it converts the received RF signal to some analog representation, typically at a much lower IF or in baseband.
Finally, the antenna stage (ANT) is the interface between the wireless channel and the RF front end.


\section{Subsystems of Radar Kit in Detail}
As illustrated in fig. \ref{fig:radar_system_top_level_architecture} b), we use COTS available FMCW radar front ends as a single building block implementation of the RF and ANT stages.
The baseband processor is realized by a RedPitaya STEMlab platform \cite{redpitaya_website}.
The students' computer and the Python programming language are typically used for the signal processing block, but signal processing could also be entirely implemented on the STEMlab \gls{soc}.
The fundamental concept of the lab radar kit is the central element of using a high-performance \gls{bbif} processor, such as the STEMlab.
This puts a stronger emphasis on the system level and processing aspects of modern automotive and industrial radar systems, which tend more and more towards software-defined radar.
Following this concept, the RF stages of our radar kit are modules added to the central processor: depending on the capabilities of different RF modules, different sensing performance can be achieved.

\subsection{Baseband \& IF Section}
\begin{figure}
    \scriptsize
    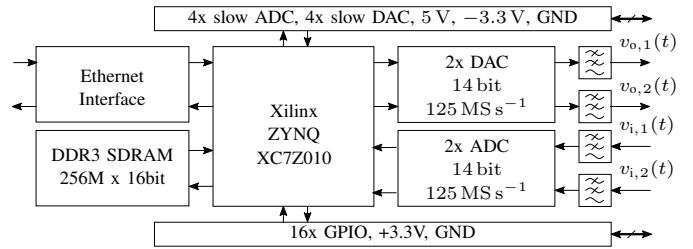
    \caption{Simplified architecture of the STEMlab with resources relevant for its application as \gls{bbif} processor. Cutoff frequency of the ADC/DAC lowpass filters is \SI{50}{\mega\hertz}.}
    \label{fig:baseband_architecture}
\end{figure}

\Glspl{soc} such as the Xilinx Zynq\textregistered-7000 family combining ARM processor cores with FPGA programmable logic have become a widespread platform for signal-processing applications, especially in combination with fully integrated transceiver ICs in the field of software-defined radio.
The combination of such \glspl{soc} with high-speed digital-to-analog converters (DACs) and analog-to-digital converters (ADCs) can yield extremely versatile baseband processing platforms.
One example is the STEMlab platform from RedPitaya \cite{redpitaya_website}.
Its functional architecture is illustrated in figure \ref{fig:baseband_architecture}.
With two DACs and two ADCs offering a sample rate of \SI{125}{\mega\sample\per\second} at \SI{14}{\bit} of resolution, together with several general-purpose input-output (GPIO) ports, auxiliary slow ADCs, DACs, and resources for external clock synchronization it is a versatile platform for realizing \gls{bbif} processing and has already been used in radar research platforms \cite{jordan2019}.
The STEMlab is connected via a gigabit Ethernet interface or via a Wi-Fi dongle.
This network connection allows to access several apps running on the Linux operating system on the ARM processor cores.
In the current version of the radar lab kit, the STEMlab is handled as a remote data generation/acquisition platform and accessed via Standard Commands for Programmable Instruments (SCPI).
Besides its use as a baseband processor, it comes with some ready-to-use measurement applications and, e.g., can also be used as an oscilloscope or arbitrary waveform generator.
It hence also provides a baseline of measurement equipment to the students.

The STEMlab is used to generate all necessary control signaling for the radar front ends.
In particular those include the control voltages for the VCOs, generated via one of the STEMlab high-speed DACs.
The two ADCs are used to acquire the IF signals from the radar RF front end.
Those can be either inphase (I) and quadrature (Q) components in case of a front end with IQ demodulator, or the signals obtained from two independent receive channels with real-valued mixers, to allow for direction-of-arrival estimation.
The ADCs can also be used to acquire the RF waveform output by the front end via a prescaler, and hence, e.g., linearity of the FMCW chirps can be analyzed.

\subsection{RF Front End}
\begin{figure}
    \scriptsize
    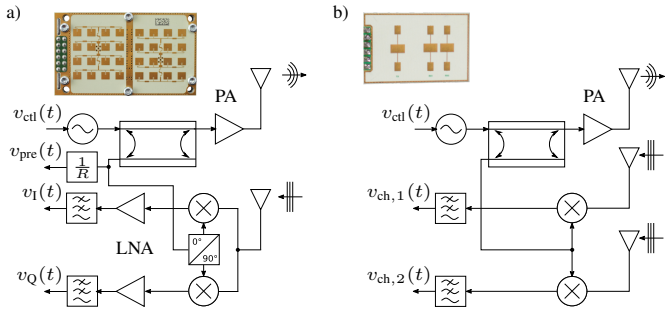
    \caption{Simplified architecture of the RF front ends used. a) InnoSenT IVS-947 and b) InnoSenT IVS-585.}
    \label{fig:frontend_architecture}
\end{figure}

COTS fully-integrated radar front ends are available at low cost, even in low volumes.
In this initial design of the radar lab kit, we selected the IVS-947 and IVS-565 \SI{24}{\giga\hertz} CW/FMCW/FSK radar front ends from InnoSenT \cite{innosent_website}, supporting a bandwidth of up to $W=\SI{250}{\mega\hertz}$.
Their basic architecture is illustrated in figure \ref{fig:frontend_architecture}.
The IVS-947 contains a single receive channel with IQ mixer.
Hence, it allows for evaluation of the phase of the Rx signal, particularly the discrimination between positive and negative Doppler shifts when used in CW mode.
Due to an integrated low-noise amplifier, the front end can directly be connected to the baseband processor without any additional circuitry and hence allows for the most simple realization of a radar system.
A prescaler divides the VCO signal by a factor of $R=8192$ and outputs it on a separate pin.
The IVS-565, in contrast, uses two Rx channels with a real-valued mixer each, connected to two independent Rx antennas.
This indeed reduces the signal information available in each channel, but enables the realization of simple direction-of-arrival estimation techniques.
Since the IVS-565 lacks an embedded LNA, it requires some external circuitry to amplify the output signal to a voltage range suitable for the baseband processor.
It cannot be directly connected to the STEMlab.
Both front ends are powered from the \SI{+3.3}{\volt} and \SI{+5}{\volt} pins of the STEMlab.
No additional power supply is necessary.

\subsection{Signal Processing}
The Python programming language is chosen for signal processing since it is open-source, multi-platform, and a universal programming language attractive for engineering education \cite{barba2020}.
The Jupyter Lab web-based user interface/integrated development environment is used to provide all lab content in a well-illustrated and documented way together with interactive data plots.
The whole radar system can be accessed and controlled via self-contained Jupyter notebooks, each notebook covering one specific topic of the radar course using the kit.


\subsection{Interfaces and Auxiliary Equipment}
\begin{figure}
    \scriptsize
    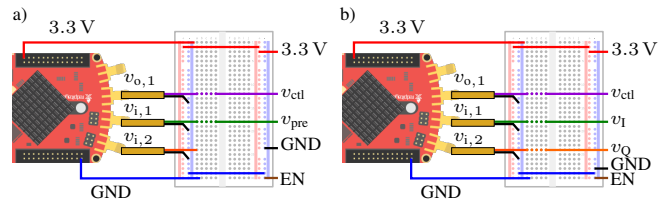
    \caption{Connections between \gls{bbif} stage and IVS-947 front end for a) chirp characterization and b) FMCW measurements.}
    \label{fig:experimental_connections}
\end{figure}

The radar front ends utilize 1x6 and 2x6 pin male headers for connection.
Adapter cables provided with the lab kit allow for interfacing the radar front ends via the corresponding female sockets to a bundle of color-coded jumper cables (see fig. \ref{fig:lab_kit_photo}).
The baseband processor GPIO, power supply and auxiliary outputs are connected with jumper cables.
The high-speed DAC and ADC paths adapters from SMA to jumper wires are used.
The jumper wires from the front ends and the wires from the \gls{bbif} stage are then interconnected using a standard breadboard.
The course materials illustrate the correct configuration using the standard color coding of the cables (see figure \ref{fig:experimental_connections}).
This approach allows for a flexible re-routing of the signals for various experiments while minimizing the risk of incorrect operation/cabling.
Additional components such as e.g. LNAs should be used in the future on the breadboard for further improving the performance and the selection of experiments.
The system is typically connected to the student’s computer via an Ethernet cable for data transfer and a USB cable for power supply.
It is completely independent of any other equipment and transportable.
This allows for, e.g., conducting outdoor measurements. 
The radar system could even be physically detached from the computer, further increasing its mobility, using a Wi-Fi connection and a USB power bank.

\section{Radar Experiments}
The use of the platform is outlined here with two experiments, conducted similar by the course attendees.

\subsection{Chirp Characterization}
\begin{figure}
    \includegraphics[]{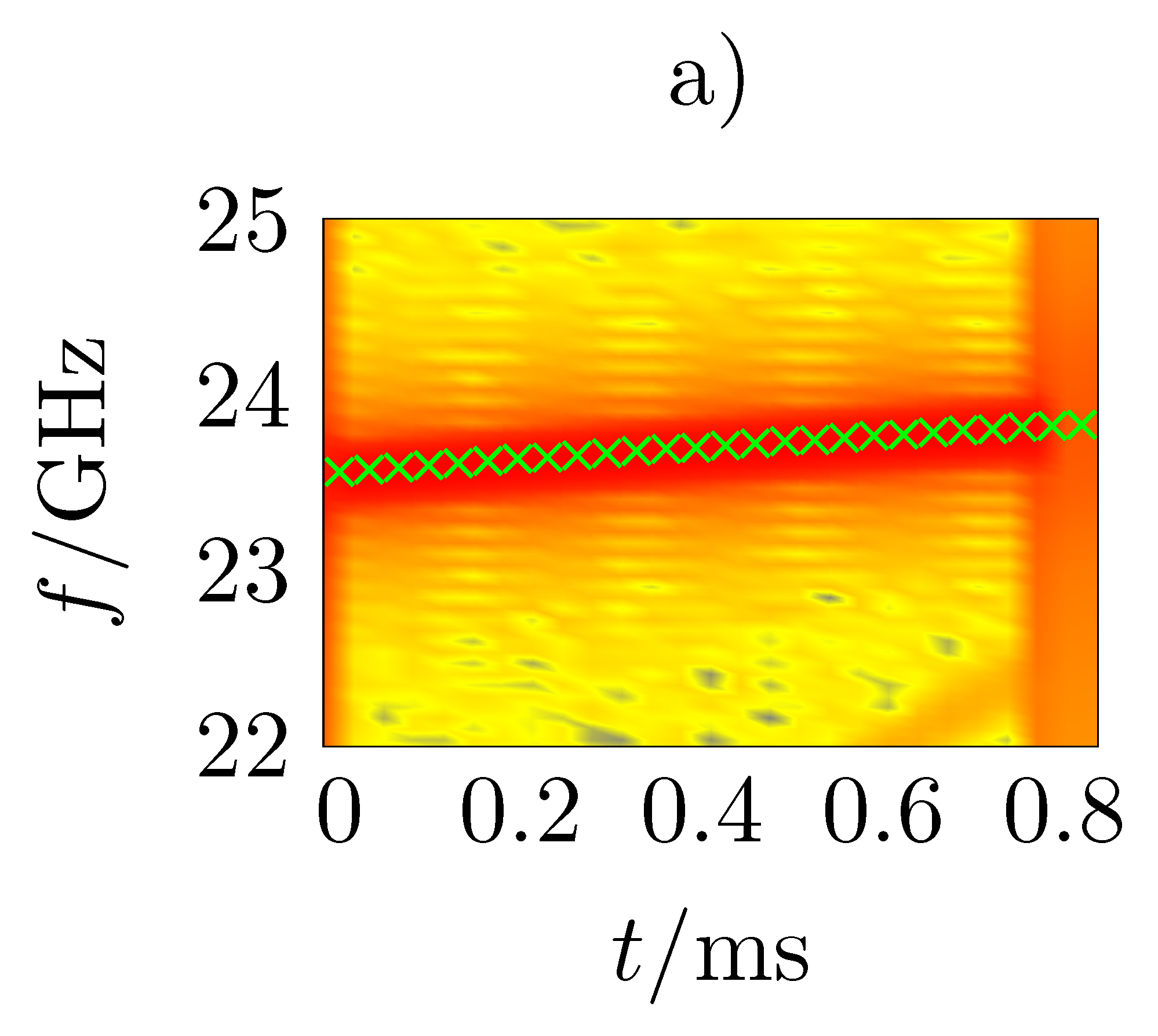}
    \includegraphics[]{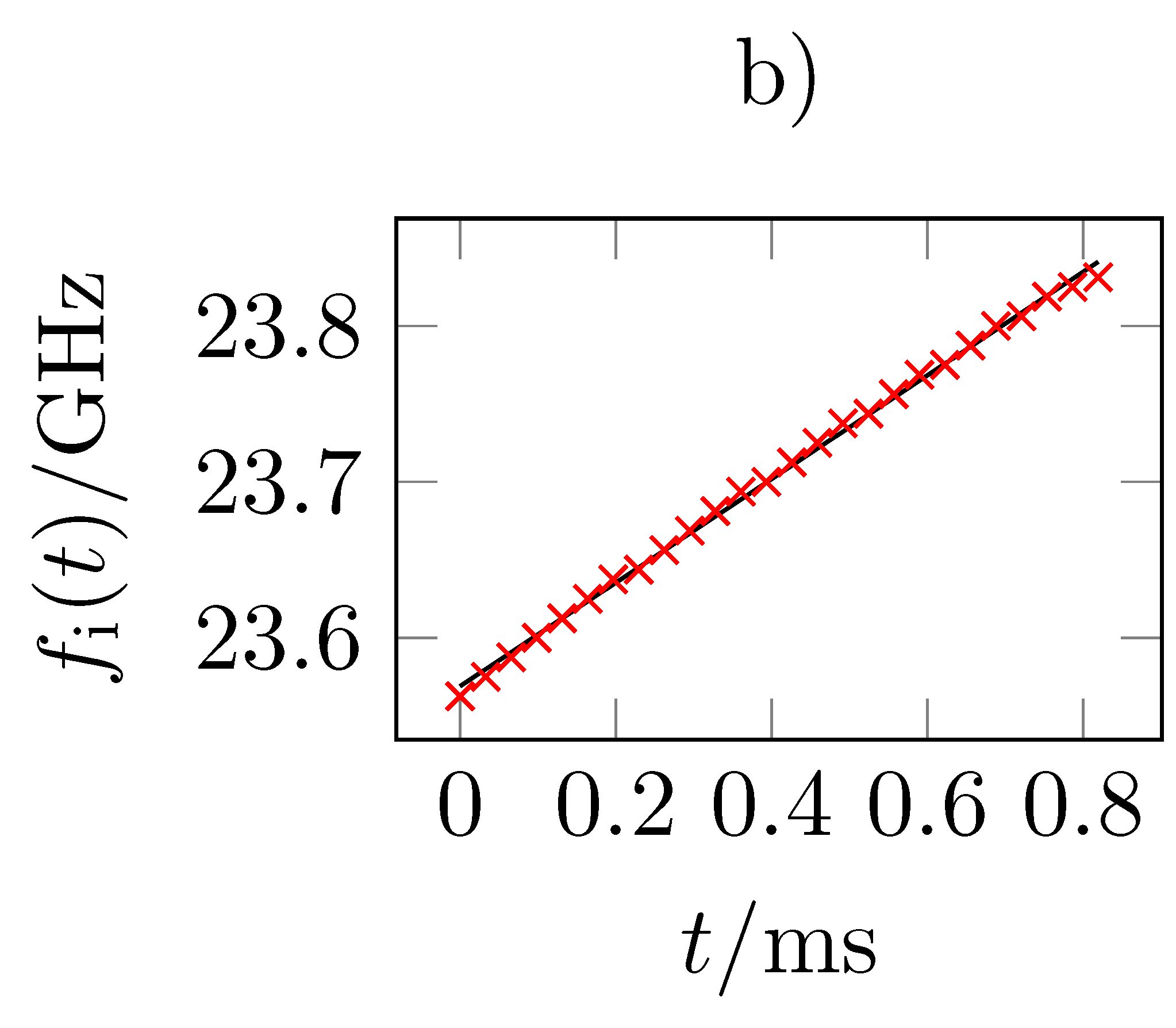}
    \caption{Example experiment A: Ramp linearity analysis. Left: STFT of acquired prescaler output and detected peaks (green crosses) and right: extracted instantaneous frequency measurements and linear regression line.}
    \label{fig:example_experiment_chirp_analysis}
\end{figure}

The chirps generated by the fronted can be analyzed by connecting the prescaler output $v_{\text{pre}}$ of the IVS-947 to $v_{\text{i},1}$ of the \gls{bbif} stage, compare fig. \ref{fig:experimental_connections} a).
Then a single upsweep control voltage ramp, increasing from \SI{0.7}{\volt} to \SI{1}{\volt} in $T=\SI{800}{\micro\second}$ is generated.
Based on its typical VCO sensitivity of $k_{\text{VCO}} = \SI{720}{\mega\hertz\per\volt}$ a linear FM chirp of $W = \SI{216}{\mega\hertz}$ is expected.
Synchronous to control voltage generation, $v_{\text{i},1}$ is sampled at a rate of $f_{\text{s}} = \SI{15.625}{\mega\sample\per\second}$ into the Rx buffer of size $N = 16384$, corresponding to an observation time of $T \approx \SI{1}{\milli\second}$.
A Short-Time Fourier Transform (STFT) with a window length of $L=256$ is applied (fig. \ref{fig:example_experiment_chirp_analysis} a), and the instantaneous frequency $f_{\text{i}}(t)$ of the chirp is extracted via peak-detection along each column of the STFT and a subsequent quadratic interpolation using the left and right neighbors of the peak.
A linear ramp can be fitted on the measured instantaneous frequency points using linear regression.
The deviation of each individually measured frequency point can be used to analyze deviations from the ideal linear slope.

\subsection{Chirp-Sequence Measurements}

\begin{figure}
    \includegraphics[]{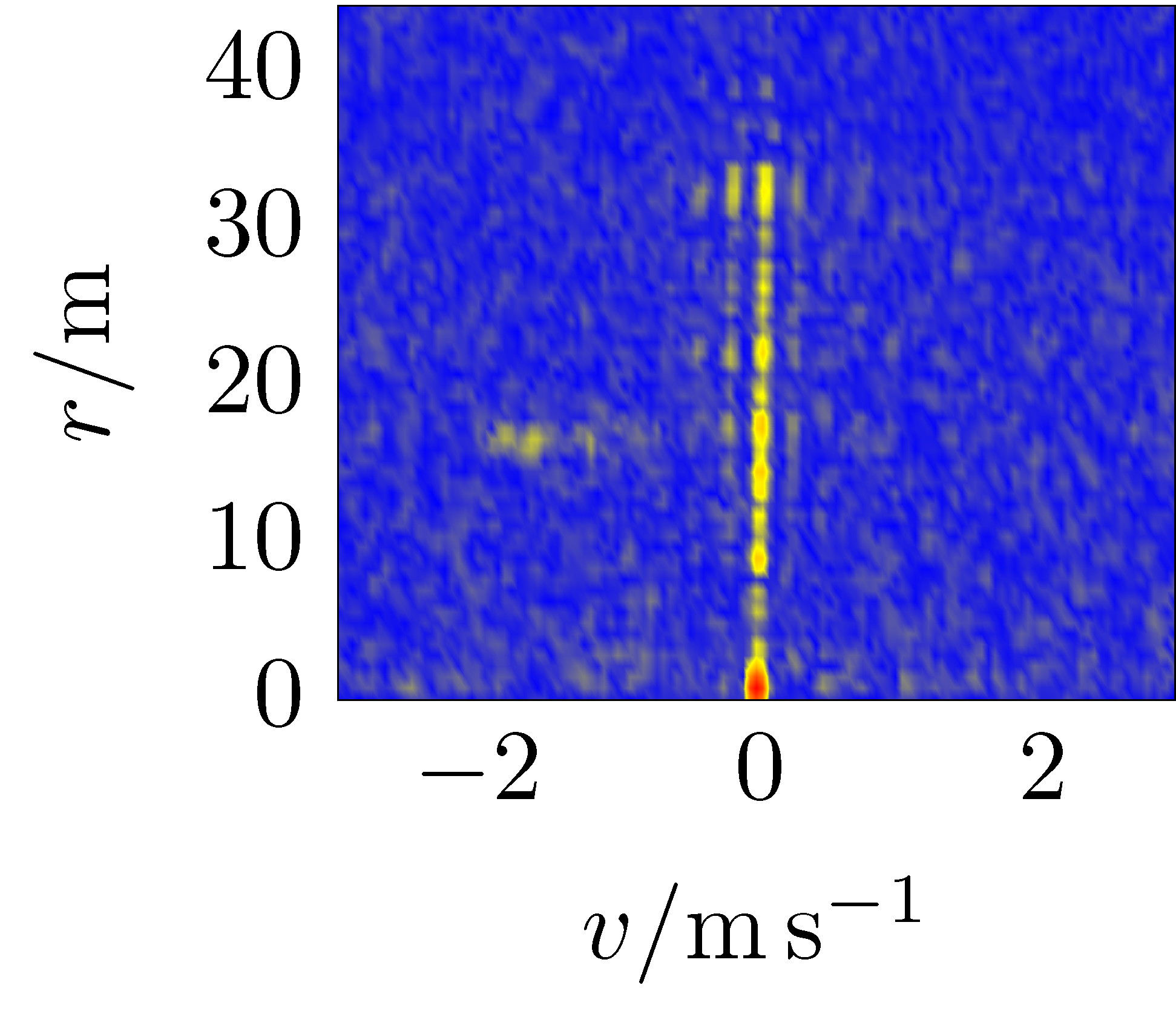}
    \raisebox{6mm}{
\begingroup%
  \makeatletter%
  \providecommand\color[2][]{%
    \errmessage{(Inkscape) Color is used for the text in Inkscape, but the package 'color.sty' is not loaded}%
    \renewcommand\color[2][]{}%
  }%
  \providecommand\transparent[1]{%
    \errmessage{(Inkscape) Transparency is used (non-zero) for the text in Inkscape, but the package 'transparent.sty' is not loaded}%
    \renewcommand\transparent[1]{}%
  }%
  \providecommand\rotatebox[2]{#2}%
  \newcommand*\fsize{\dimexpr\f@size pt\relax}%
  \newcommand*\lineheight[1]{\fontsize{\fsize}{#1\fsize}\selectfont}%
  \ifx\svgwidth\undefined%
    \setlength{\unitlength}{99.21259843bp}%
    \ifx\svgscale\undefined%
      \relax%
    \else%
      \setlength{\unitlength}{\unitlength * \real{\svgscale}}%
    \fi%
  \else%
    \setlength{\unitlength}{\svgwidth}%
  \fi%
  \global\let\svgwidth\undefined%
  \global\let\svgscale\undefined%
  \makeatother%
  \begin{picture}(1,0.85714286)%
    \lineheight{1}%
    \setlength\tabcolsep{0pt}%
    \put(0,0){\includegraphics[width=\unitlength,page=1]{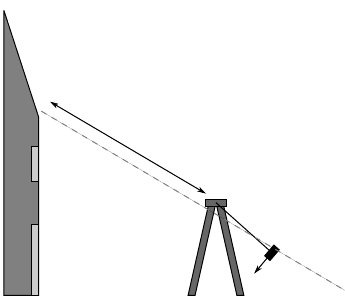}}%
    \put(0.39253626,0.44373121){\color[rgb]{0,0,0}\makebox(0,0)[lt]{\lineheight{1.25}\smash{\begin{tabular}[t]{l}$r$\end{tabular}}}}%
    \put(0,0){\includegraphics[width=\unitlength,page=2]{fmcw_scenario.pdf}}%
  \end{picture}%
\endgroup%
}    
    \caption{Example experiment B: Chirp-sequence measurement from knee wall of a house's roof down into garden, pointing towards a moving child swing.
    The swing is clearly visible at a range of $r\approx\SI{15}{\meter}$ and a radial velocity of $v\approx\SI{-2}{\meter\per\second}$.}
    \label{fig:example_experiment_fmcw_measurement}
\end{figure}

For chirp-sequence radar measurements, the I and Q outputs of the IVS-947, $v_{\text{I}}$ and $v_{\text{Q}}$, are connected to both ADC inputs $v_{\text{i},1}$, $v_{\text{i},2}$, compare fig. \ref{fig:experimental_connections} b).
The waveform generator of the STEMlab then creates a control voltage signal for $N=128$ upchirps of bandwidth $W = \SI{216}{\mega\hertz}$, each of duration $T_{\text{c}} \approx \SI{1}{\milli\second}$.
Synchronous sampling of the inputs is performed at a rate of $f_{\text{s}} = \SI{122.07}{\kilo\sample\per\second}$ and the vector of acquired samples is transformed into a fast-time/slow-time matrix of shape $126 \times 128$.
A 2D Fourier transform yields the Range-Doppler matrix as e.g. illustrated in fig. \ref{fig:example_experiment_fmcw_measurement} for an exemplary scene.

\section{Summary, Limitations, and Next Steps}

We have introduced a radar lab kit for hands-on distance learning consisting of just two COTS low-cost subsystems: a baseband processor utilizing a modern \gls{soc} and high-speed DAC and ADC channels, and \SI{24}{\giga\hertz} CW/FMCW/FSK radar front ends.
Measurements from typical lab experiments were used to illustrate that the system can be used for versatile experiments in radar systems education.
High data rate waveforms such that chirp-sequence FMCW are supported.
A current limitation is the use of SCPI commands for data transfer, increasing the minimum measurement repetition rate to \SI{450}{\milli\second}.
In addition, the use of FPGA block-ram for waveform storage is limited to 16384 samples for each, Tx and Rx, requiring a careful trade-off between sampling rate and maximum available continuous observation time.
Both can be addressed by software and FPGA firmware modifications, e.g., utilizing the DDR3 SDRAM for data storage.
The radar kit in its current state offers an excellent solution for practical radar education and has considerable potential for future improvements.
A course based on this kit was first conducted in the winter semester of 2020.
A systematic evaluation at the end of the course resulted in an excellent mean rating of 1.2 on a scale from 1 to 6.


\bibliographystyle{IEEEtran}

\bibliography{IEEEabrv,IEEEexample,paper_eumw_2021}

\end{document}